\documentclass{appolb}
\usepackage{graphicx}
\usepackage{amssymb}

\begin{document}
\title{Exploring high-multiplicity events in high-energy proton-proton collisions%
\thanks{Presented at Diffraction and Low-x 2024}%
}
\author{Yuri N. Lima, Victor P. B. Gon\c calves
\address{Universidade Federal de Pelotas}
\\[3mm]
{Andr\'e V. Giannini 
\address{Universidade Federal da Grande Dourados}
}
}
\maketitle
\begin{abstract}
It is known that the proton is overpopulated by gluons and is characterized as a highly dense medium at high collision energies. From this, the formation of a new state of matter called Color Glass Condensate (CGC) is expected, and an open question is whether the nonlinear effects predicted by this state are identifiable at the LHC. The multiplicity of particles produced in a hadronic collision presents as a means to adequately investigate this problem. Currently, the description of the available data for different multiplicity regimes remains a challenge. Even though different experimental collaborations have identified that the production of certain final states, in $pp$ collisions, present a modification in the behavior of high multiplicity events in relation to the case of minimum bias we still lack a way to identify the nature of those high multiplicity events: are those driven by initial-state effects, final-state effects or a mix of both? We argue that a analyzing different particle production process that can be described CGC framework, in particular, isolated photon production which is not sensitive to final-state effects, may provide a path forward in answering this question.
\end{abstract}
  
\section{Introduction}

Understanding processes described by Quantum Chromodynamics (QCD) is an important objective of Particle Physics. Some of these processes present open questions, such as, what is the nature of high multiplicity events? From the initial-state point of view, while events with multiplicities close to the minimum bias one may be related to typical color charge configurations of the proton, high-multiplicity events may be correlated with rare color charge configurations. The origin of such events is still in debate, as there is also the possibility that rare events may be driven by final state effects or a mixture of initial- and final-state effects.

Presently, it is well established that the production of different final states versus co-produced charged particles has a behavior at high multiplicities different from the minimum bias case, displaying a non-linear dependence as multiplicity increases~\cite{ALICE:2019avo,ALICE:2020msa}. Even though we still lack a quantitative description of the available data using a single theoretical framework, here we present a systematic investigation of the correlation the multiplicity of different final-states as a function of the multiplicity of co-produced charged particles. employing the ``hybrid formalism''~\cite{Dumitru:2005gt} of the CGC effective field theory, which describe the particle production as a convolution of collinear and transverse momentum dependent distributions together with a fragmentation function, responsible for converting partons to final-state hadrons. Following previous studies of the high multiplicity events using the CGC formalism~\cite{Ma:2018bax,Levin:2019fvb,Kopeliovich:2019phc,Gotsman:2020ubn,Siddikov:2020lnq,Siddikov:2021cgd,Stebel:2021bbn,Salazar:2021mpv}, we will assume that the particle production mechanism is the same for low and high multiplicity events, with the main difference being the saturation scale, $Q_s$.

Given the ``dilute-dense'' configuration assumed in the hybrid formalism, where saturation effects are only accounted for in the target, we can anticipate that collisions with the highest multiplicities at central rapidity will not be well described. For this reason, we only present results for particle production at forward rapidities.

\section{Brief review of formalism}

\begin{figure}[htb]
\centerline{
\includegraphics[width=12.5cm]{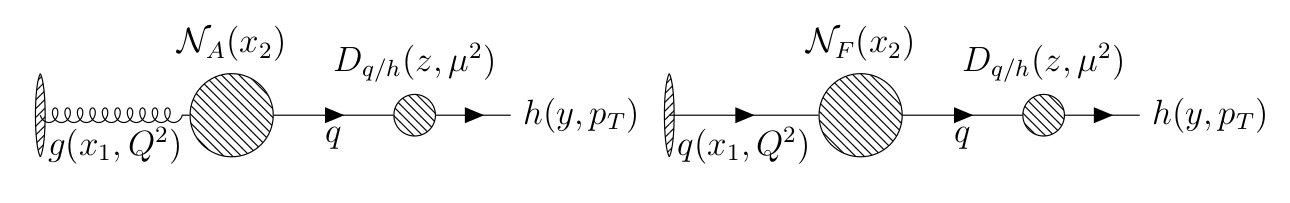}}
\caption{Schematic representation of a heavy meson production in the hybrid formalism: a gluon (left diagram) and a quark (right diagram) of the incident hadron interact with the target and subsequently hadronize into the heavy meson.}
\label{hybridformalismdiagram}
\end{figure}

Given the ``dilute-dense'' configuration assumed in the hybrid formalism the cross section for producing a hadron with transverse momentum $p_T$ at a given rapidity $y$ can be expressed as the following convolution (see Fig. \ref{hybridformalismdiagram}):
\begin{equation}
\sigma(pp\rightarrow hX)\propto g(x_1,Q^2)\otimes\mathcal{N}_A(x_2)\otimes D_{q/h}
+
\sum_i q_i(x_1,Q^2)\otimes\mathcal{N}_F(x_2)\otimes D_{q_i/h}\,,
\label{Eq_sig_imp}
\end{equation}
where $g(x_1,Q^2)$ and $q_i(x_1,Q^2)$ ($i = c$ for $D^0$-meson and $i = u, d, s$ for light mesons) are the gluon and quark densities, respectively, of the projectile, calculated at $x_{1,2}=(p_T/\sqrt{s})e^{\pm y}$; $\mathcal{N}_{A,F}$ are, respectively, the adjoint and fundamental scattering amplitudes, which encode all the information about the scattering and, therefore, about the nonlinear and quantum effects on the hadron wave function. These scattering amplitudes will be solutions of the Balitsky-Kovchegov equation for different initial conditions. $D_{g/h}$ ($D_{q/h}$) is the fragmentation function of a gluon $g$ (quark $q$) into a hadron $h$. We refer to~\cite{Lima:2022mol,Lima:2024ksd} for the full expressions.

The first term of the above equation refers to the gluon-initiated subprocess, which is expected to be dominant at high energies and non-forward rapidities; the second contribution is due to quark-initiated processes which becomes the dominant contribution at (ultra-)forward rapidities, providing a way to study non-perturbative fluctuations of the hadronic wave function in the form of a intrinsic heavy-quark component. Such a possibility was also considered in ref.~\cite{Lima:2024ksd}.

In the next section, in addition to results for $K^0_S$ and $D_0$ mesons, we also present results for isolated photon production in high-energy pp collisions. In the same fashion as Eq. \ref{Eq_sig_imp}, the photon production cross-section from the $q+g$ scattering can be written as
\begin{equation}
	\sigma(pp\rightarrow \gamma X)\propto \sum_i q_i(x_1,Q^2)\otimes\mathcal{N}_F(x_2)\,,
	\label{Eq_sig_photons}
\end{equation}
where $i = u, d, s$. We refer to~\cite{Lima:2023dqw, Ducloue:2017kkq} for the full expression and kinematics. It is important to note that the photon production is not sensitive to final-state effects, and when associated with other (hadronic) final states provides a way to improve our knowledge about high multiplicity events as well as the limitations of the current formalism employed in our calculations.

\section{Results and discussion}

\begin{figure}[htb]
\centerline{
\includegraphics[width=12.5cm]{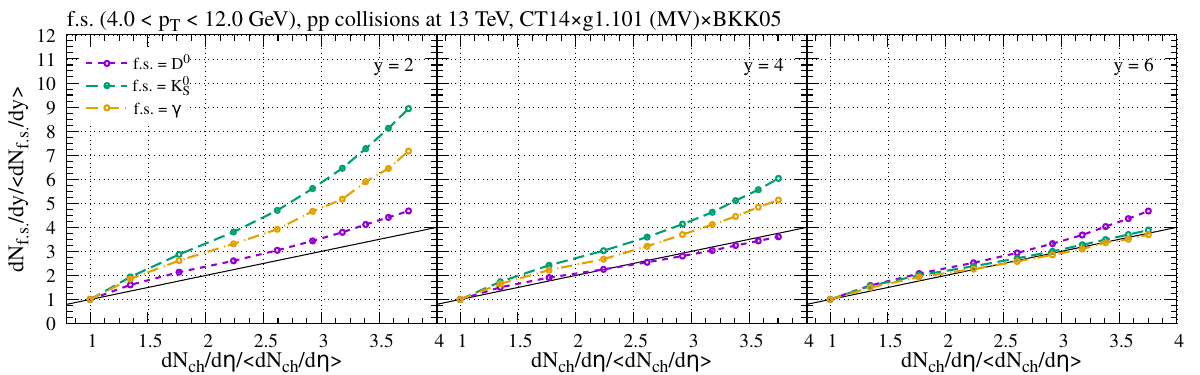}}
\caption{Correlation of the normalized multiplicity of $K^0_s$ and $D^0$ mesons and $\gamma$ as a function of the normalized multiplicity of charged hadrons produced in $pp$ collisions at 13 TeV and different rapidity values.}
\label{correlations}
\end{figure}

Figure \ref{correlations} shows the predictions for the normalized yield of three different final-states for different rapidities ($y = 2, 4, 6$) in the range $4\le p_T\le 12$ GeV as a function of the co-produced charged particles produced in the central pseudo-rapidity region ($|\eta|<0.5$). Here we used the CT14 parameterization \cite{Dulat:2015mca} for the parton distribution functions, BKK05~\cite{Kniehl:2006mw} (AKK08~\cite{Albino:2008fy}) for the fragmentation functions of heavy (light) mesons and a solution of the BK equation based on the MV model from the AAMQS fits~\cite{Albacete:2012xq}.

One can see that for $2 < y \lesssim 4$ the behavior of the correlation of normalized multiplicities is final-state dependent, with different magnitudes depending on the type of particles observed. For larger rapidities, the result becomes final-state independent, with the three curves nearly collapsing on top of each other. Such a result is expected in an approach where the nature of high multiplicity events is solely due to initial-state effects, as rare color charge fluctuations: for large rapidity values $Q_s$ becomes the dominant momentum scale, thus saturation effects affect high $p_T$ particles in the same way it affects low $p_T$ ones. In particular, as the photon production cross-section does not depend on final-state effects, its measurement for different rapidity values may allow further tests on the current limitations of the initial-state effects as described by the CGC formalism, which also impacts hadronic final-states.

In conclusion, a systematic study of the multiplicity of different final-states employing the hybrid formalism of the CGC effective field theory was conducted. The present predictions motivate measurements of the produced yield of different final-states in different rapidity regions as a possible mean to advance our understanding of high multiplicity events in high energy proton-proton collisions.

\section*{Acknowledgements}
This work was partially supported by INCT-FNA (Process No. 464898/2014-5). V.P.G. was partially supported by CNPq, CAPES and FAPERGS. Y.N.L. was partially financed by CAPES (process 001).  The authors acknowledge the National Laboratory for Scientific Computing (LNCC/MCTI, Brazil), through the ambassador program (UFGD), subproject FCNAE for providing HPC resources of the SDumont supercomputer.

\end{document}